# Generation of coherent phonons in bismuth by ultrashort laser pulses in the visible and NIR: displacive versus impulsive excitation mechanism.

A.A. Melnikov[1,2], O.V. Misochko[3], and S.V. Chekalin[1]

[1]*Institute of Spectroscopy, Russian Academy of Sciences, 142190 Troitsk, Moscow region, Russia*
[2]*Moscow Institute of Physics and Technology, 141700 Dolgoprudny, Moscow region, Russia*
[3]*Institute of Solid State Physics, Russian Academy of Sciences, 142432 Chernogolovka, Moscow region, Russia*

**Abstract**
We have applied femtosecond pump-probe technique with variable pump wavelength to study coherent lattice dynamics in Bi single crystal. Comparison of the coherent amplitude as a function of pump photon energy for two different in symmetry $E_g$ and $A_{1g}$ phonon modes with respective spontaneous resonance Raman profiles reveals that their generation mechanisms are quite distinct. We show that displacive excitation, which is the main mechanism for the generation of coherent $A_{1g}$ phonons, cannot be reduced to the Raman scattering responsible for the generation of lower symmetry coherent lattice modes.

**1. Introduction**

When ultrashort laser pulses interact with solids they can excite lattice vibrations showing high degree of temporal and spatial coherence [1-3]. These coherent phonons reveal themselves as damped oscillations of a certain optical parameter (e.g. reflectivity or transmittance) in the time domain. To create coherence one needs to establish fixed phase relations either among different q modes of the same phonon branch [4], or between vacuum and excited states of a single $q \approx 0$ phonon mode [5]. Coupling of these lattice states can be achieved in two distinct ways. First, owing to the large spectral bandwidth of an ultrashort laser pulse, it can generate nonstationary phonon states in a crystal through impulsive stimulated Raman scattering, providing a kick to the atoms of the crystal [6, 7]. In this case, which we will further refer to as Raman mechanism (RM), the coherence is "field driven", with the driving force responsible for both the energy and coherence transfer from electromagnetic field to lattice. Apart from the "field driven" coherence directly prepared by laser radiation, lattice coherence can also be driven by rapid nonradiative processes. For opaque crystals that have long-lived excited states, the excited state coherence is usually dominant [1-3]. To describe the creation of coherence in this case the displacive excitation of coherent phonons (DECP) model was introduced [8]. In contrast to RM, for DECP the lattice coherence is "relaxation-driven" with no momentum imparted to the atoms from the pump pulse. Additionally, DECP can generate only fully symmetric coherent phonons, whereas RM is capable to create coherent phonons of full and lower symmetry, provided the latter belong to even (gerade) representations of a space group. The primary difference is that in DECP selection rules for the pump play no role and coherent amplitude is determined by the absorption coefficient alone, while in RM the selection rules are controlled by the symmetry of Raman tensor. Thus, DECP is controlled by first order (in electromagnetic field) process, but RM is governed by a second order process.

In an attempt to describe the creation of lattice coherence within a unified approach it was suggested that DECP is not a distinct mechanism, but a particular case of RM [9, 10]. Specifically, the off-resonance excitation by laser light pulses at wavelengths lying in the transparency region of a crystal was predicted to impose impulsive driving force on atoms resulting in sine-like coherent oscillations. On the other hand, when the pump radiation falls into resonance with electronic transitions the excitation should be displacive and the atomic motion is then characterized by a cosine-like phase. Each of the two limiting cases is described by a component of a Raman tensor introduced in [10].

However, in many crystals an intermediate initial phase of coherent phonons is observed and therefore, to bring this Raman-based model into agreement with experimental data, its further elaboration included the effects of finite lifetime of excited charge density on the phase and amplitude of coherent phonons [11]. The concept is based on the fact that the decay time of charge carrier density defines the duration of driving force acting on atoms. Provided the decay is rather fast (as compared to the oscillation period), the excitation of coherent phonons acquires impulsive features even in the resonant case. Thus, in time domain the DECP model loses its distinctive features and seemingly should be replaced by a unified RM [11].

One straightforward way to check the validity of the generalized Raman model is to measure the initial phase of oscillations in the photoinduced response. However, this approach has at least two disadvantages. The first is that for a number of semimetals (Te, Bi, and Sb) DECP and RM give nearly identical predictions for the initial phase of fully symmetric phonon modes [11]. The second results from the difficulty in a precise phase determination. The latter procedure deals with a signal at the moment of the pump and probe pulses overlap and thus it is usually ambiguous.

In the present study an alternative approach is used. We have measured the dependence of coherent phonon amplitude of fully symmetric $A_{1g}$ and doubly degenerate $E_g$ phonon modes in bismuth on excitation wavelength in visible and near infrared ranges. According to the unified Raman model, the coherent phonon amplitudes should be proportional to the corresponding Raman cross sections. Moreover, at any given wavelength the relative intensities of excited phonon modes obtained from the frequency- and time-domain data should be identical. Yet, a comparison of our pump-probe results with spontaneous Raman measurements obviously contradicts the predictions of the unified Raman theory, showing that DECP cannot be reduced to RM.

**2. Experimental details**

In our experiments all measurements were done at room temperature with a typical ultrafast pump-probe setup. Amplified pulses of a Ti:Sapphire laser operating at $\lambda$=800 nm were divided into two parts. The first was attenuated to be used to probe the sample – a single crystal of bismuth oriented in such a way that its surface was perpendicular to the trigonal axis. The second part of the beam was used to seed a parametric amplifier to provide pump pulses of 70 fs duration with a tunable central wavelength. The pump and probe beams were incident nearly perpendicular to the surface of the crystal. We detected a component of the probe beam reflected from the crystal and polarized parallel to the pump polarization, while the probe pulses were polarized at 45° relative to the pump. Energy of this component was measured with opened and closed pump beam and the relative change in reflectivity $\Delta R/R_0$ was recorded at a given time delay between the pump and probe pulses. The size of the probe beam spot was about 50 μm in all experiments. The size of the pump beam had a minimum value of about 120 μm for 400 nm pulses and increased at longer pump wavelengths (e.g. for 1300 nm the spot size was as large as 400 μm).

**3. Results**

Figure 1 shows the transient reflectivity obtained with pump wavelengths of $\lambda$=400 and $\lambda$=1300 nm. Ultrafast excitation at t=0 is followed by a monotonic decay on which pronounced oscillations are superimposed. Since the dominant oscillation frequency almost matches that of $A_{1g}$ phonon mode, these oscillations are the result of fully symmetric $A_{1g}$ coherent atomic motion induced by the laser pulse. It should be emphasized that we did not define zero time delay and therefore, the initial phase of coherent oscillations remains unknown. Nevertheless, in the previous time-domain studies [8, 12] the phase of $A_{1g}$ mode was measured for 600 and 800 nm excitation wavelengths and we use this result here, assuming $A_{1g}$ oscillations to be cosine ones, at least in the visible excitation range.

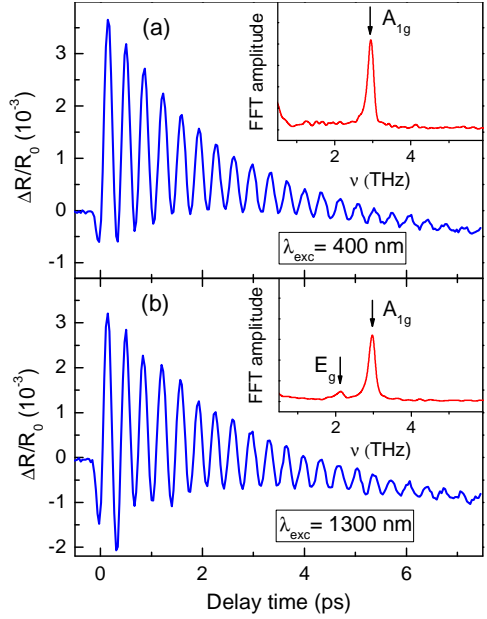

Fig. 1. (a) Transient reflectivity of bismuth excited at $\lambda$=400 nm as a function of time delay. The inset shows normalized Fourier spectrum for the oscillatory part of the signal. (b) The same curves for the case of $\lambda$=1300 nm excitation. Note that in the inset apart from the $A_{1g}$ mode, the $E_g$ mode is clearly visible.

Taking into account the above, the transient reflectivity excited by the $\lambda$=400 nm pump pulses can be well fitted at positive time delays by the sum of a damped cosine with varying frequency and a monotonic term:

$\Delta R/R_0 = A(A_{1g})exp(-\gamma_1 t)cos(\nu_1(t)t) + B(t)$  (1)

where $A(A_{1g})$ is the amplitude and $\gamma_1$ - is the decay rate of coherent $A_{1g}$ oscillations. The exact form of $B(t)$ term is of no importance in the present study; however, we note that the term has a multi-exponential character. The temporal dependence of frequency $\nu_1$ reflects the effect of so-called frequency or bond softening [13]. Briefly, the instantaneous frequency of $A_{1g}$ coherent phonons in bismuth is not constant due to the presence of excited charge carriers. Its decrease is the most appreciable during few initial cycles of vibrations and then the frequency gradually returns to the unperturbed value of 2.93 THz. It is reasonable in our case to treat the relaxation of the frequency shift as exponential with decay time of ~ 1 ps, while the minimal value of $\nu_1$ attained near zero time delay is ~ 2.8 THz. The photoinduced response described by Eq. (1) is typical for bismuth and it has been observed in numerous previous experiments [8, 12, 13].

As far as the photoinduced response measured with 1300 nm pump pulses is concerned, it shows a considerable deviation from that excited by the shorter wavelength pulses, especially during the first few picoseconds. Its Fourier spectrum shown in the inset indicates that the main difference is due to the presence of strongly damped oscillations at ~ 2.1 THz, which can be naturally ascribed to the coherent phonons of $E_g$ symmetry [14, 15]. By including $E_g$ oscillations into analysis it is possible to fit the measured decay traces. The additional term is a damped harmonic oscillatory function at frequency of 2.1 THz and the only parameter to be addressed specially is its phase. As it has been already noted, we did not measure the absolute initial phase, and in the case of $A_{1g}$ mode just relied on the data available. For $E_g$ mode in bismuth the coherent oscillations reported to have sine-like pattern, with the initial phase being independent on temperature [16]. In our measurements, due to significant difference of $A_{1g}$ and $E_g$ frequencies the *relative* phase may be defined with a satisfactory precision. Therefore, we represented the overall time resolved signal in the following form:

$$\Delta R/R_0 = A(A_{1g})\exp(-\gamma_1 t)\cos(\nu_1(t)t) + A(E_g)\exp(-\gamma_2 t)\sin(\nu_2 t + \phi) + B(t) \qquad (2)$$

where $A(E_g)$, $\gamma_2$, $\nu_2$ are the amplitude, the decay rate and the frequency of $E_g$ oscillations. As a result of the fitting, for the $\lambda = 1300$ nm excitation we obtained the following oscillation parameters: $A(A_{1g}) = (1.2\pm0.1)\cdot10^{-3}$ and $A(E_g) = (3.0\pm0.5)\cdot10^{-4}$ for coherent amplitudes (the amplitude ratio of ~ 4) and $\gamma_1 = 0.5$ ps$^{-1}$ and $\gamma_2 = 1.5$ ps$^{-1}$ for the decay rates. Variation of $\phi$ gives the result shown in Fig. 2.

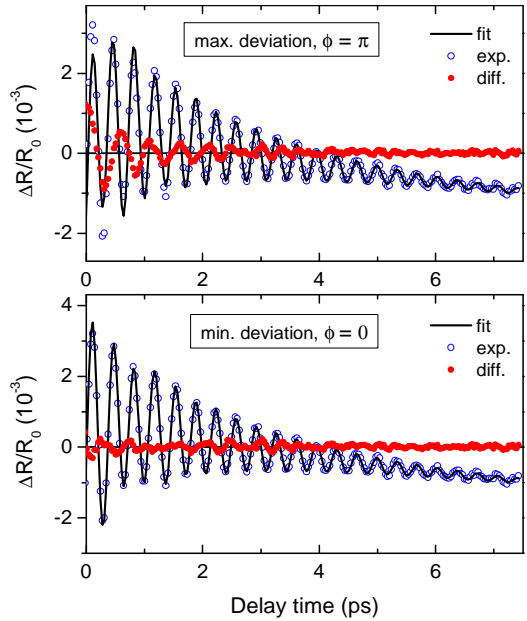

Fig. 2. Experimental data (empty circles), a fit by formula (2) (solid line) and a residual (filled circles) for two values of the sine term phase: $\phi = \pi$ and $\phi = 0$.

The minimum difference between the model signal (2) and the experimentally measured one (characterized by mean-square residual) is observed near $\phi=0$ (and consequently at multiples of $2\pi$). As $\phi$ increases the deviation also exhibits a sine-like growth, reaching its maximum value at $\phi=\pi$, see Fig. 2. Therefore, we conclude that the relative phase of $A_{1g}$ and $E_g$ oscillations is close to 90° with the error less than ±10° resulting mainly from some ambiguity in the temporal shape of $\nu_1(t)$.

To measure varying amplitudes of the coherent oscillations excited in the near infrared and visible, a set of excitation wavelengths ranging from 400 to 2500 nm was used and in each case the fitting procedure described above was repeated. The pump spot size and bismuth reflectivity were taken into account to ensure equal intensity of the pump field for all the wavelengths used (if measured inside the sample, near the surface). It was found that when the latter condition to be fulfilled, the monotonic part of the signal B(t) together with decay rates of the oscillations were almost independent of excitation wavelength. At the same time, the amplitudes of coherent phonons demonstrated significant spectral variation, essentially different for $A_{1g}$ and $E_g$ modes. Figure 3 illustrates the observed wavelength dependence of amplitude for fully symmetric and doubly degenerate modes.

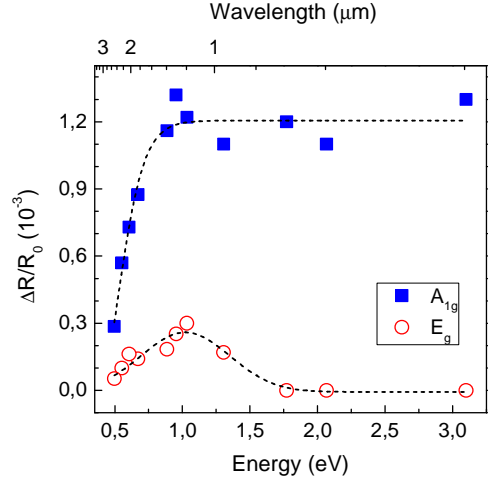

Fig. 3. Amplitudes of $E_g$ and $A_{1g}$ coherent oscillations as a function of excitation photon energy. Dashed lines are a guide to the eye.

One can see that the amplitude of $E_g$ oscillations is close to zero in the visible range for the wavelengths of $\lambda$=400, 600 and 700 nm. However, starting from $\lambda$~800 nm the $E_g$ amplitude begins to grow reaching its maximum near $\lambda$=1300 nm (~ 1 eV). In contrast, the amplitude of $A_{1g}$ phonons remains approximately constant in the visible and near infrared range up to $\lambda$=1300 nm. For excitation wavelengths larger than $\lambda$=1800 nm (~ 0.7 eV), the generation efficiency for both phonon modes exhibits a significant decrease.

**4. Discussion**

As the rhombohedral unit cell of bismuth contains two atoms, group theory predicts for Bi three $\Gamma$-point optical modes, i.e., a fully symmetric $A_{1g}$ and a doubly degenerate $E_g$ modes, both detected by spontaneous Raman scattering [15]. The atoms are displaced along the trigonal axis in the $A_{1g}$ mode, and perpendicular to that axis in the $E_g$ mode as illustrated in Fig. 4.

In time-domain, employing isotropic detection scheme, one can measure the coherent phonons having non-zero diagonal matrix elements, therefore, in bismuth both the fully symmetric $A_{1g}$ phonons with nonzero trace and non-fully symmetric $E_g$ phonons with zero trace can be created and detected. As follows from the $E_g$ Raman tensor, the detected $E_g$ amplitude should depend on the polarization of probe light. This was indeed observed not only for probe, but also for pump pulses [16]. Since the generation and detection process for $A_{1g}$ mode is polarization independent, it is possible to eliminate $A_{1g}$ contribution in anisotropic detection by subtracting signals from two channels that detect orthogonal probe polarizations [16, 17]. In our case we implemented similar configuration (see the previous section) but without full subtraction (the purpose is to increase the $E_g$ signal slightly and to ensure a more reliable measurement of its wavelength dependence).

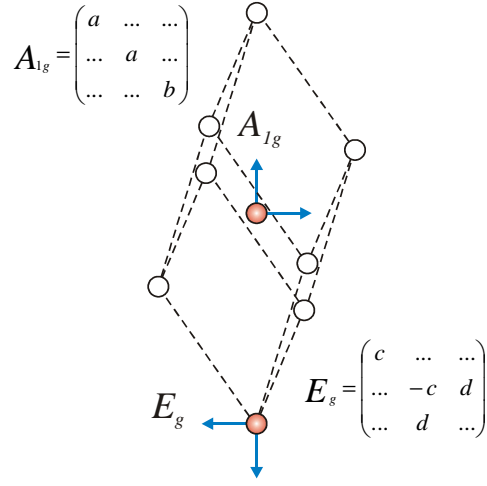

Fig. 4. A rhombohedral unit cell of bismuth with two atoms. Arrows illustrate atomic vibrations parallel ($A_{1g}$) and perpendicular ($E_g$) to the trigonal axis. Raman tensor for $A_{1g}$ and one of the two corresponding tensors for doubly degenerate $E_g$ mode are shown.

In most previous studies fully symmetric $A_{1g}$ coherent contribution dominates the transient reflectivity obtained by isotropic detection at room temperature, while doubly degenerate $E_g$ oscillations are almost negligible. The latter appear either at low temperatures [16], or at high excitation densities [14]. Alternatively, they can be observed at room temperature by a special detection technique such as anisotropic sampling [17], which relies on polarization properties of $A_{1g}$ and $E_g$ phonons. Our results clearly demonstrate that it is possible to select a particular excitation wavelength at which the amplitude of $E_g$ oscillations is relatively large even at room temperature (reaching up to one fourth of $A_{1g}$ coherent amplitude). The set of data points shown in Fig. 3 for $E_g$ phonons can be approximated by a bell-like shape, which implies that the process of coherent $E_g$ phonon generation is a resonant one at long wavelength excitation.

Here it is appropriate to refer to the existing models of the band structure of bismuth to find out which electronic transitions are responsible for the observed resonance at 1 eV. For that purpose we use the theoretical results by Golin [18] and the optical data obtained by piezoreflectance measurements [19]. It should be noted, however, that the spectrum of femtosecond pulses used in our experiments is rather broad and the same is true concerning the width of valence electronic bands in bismuth. Therefore, several potential assignments both for $E_g$ and $A_{1g}$ resonances exist. Figure 5 shows a fragment of the bismuth band structure qualitatively reproduced from [18] with pertinent optical transitions marked by arrows. Specifically, they are $L_a(5) - L_s(7)$, $\Gamma_{45}^{+}(4) - \Gamma_6^{-}(6)$ and $\Gamma_6^{+}(5) - \Gamma_6^{-}(6)$ transitions with energies equal to 1.15, 0.825 and 0.821 eV according to [18]. Piezoreflectance experiments gave slightly different values, namely 1.13, 0.81 and 0.69 eV [19]. The maximum of $E_g$ amplitude located around ~ 1 eV then can correspond to one of the mentioned optical transitions. When the carrier frequency of pump femtosecond pulse approaches this energy, the efficiency of coherent phonon generation increases strongly. If the central wavelength is detuned from resonance, the efficiency drops, making the detection of $E_g$ component by standard pump-probe technique barely possible.

According to the RM proposed by Merlin et al. [10], at resonance the phase of coherent oscillations should be cosine-like. As in bismuth the resonant conditions are fulfilled for both $A_{1g}$ and $E_g$ modes, the initial phase shift between the two modes must be close to zero. However, the analysis of our experimental data revealed the phase difference of about 90° for λ=1300 nm, as well as for every excitation wavelength at which the amplitude of $E_g$ oscillations was sizeable.

This result obviously contradicts the unified Raman model in its simple form, though the discrepancy may be in principle eliminated by taking into account the finite lifetime of excited charge carriers in a way it was done by Riffe and Sabbah [11]. To do this, one needs to assume the lifetime of excited charge density coupled to $E_g$ phonons to be much less than the pump pulse duration. Besides, the charge density should be anisotropic to satisfy symmetry conditions (which reveal themselves in experiment as the polarization dependence of $E_g$ coherent amplitude [16]).

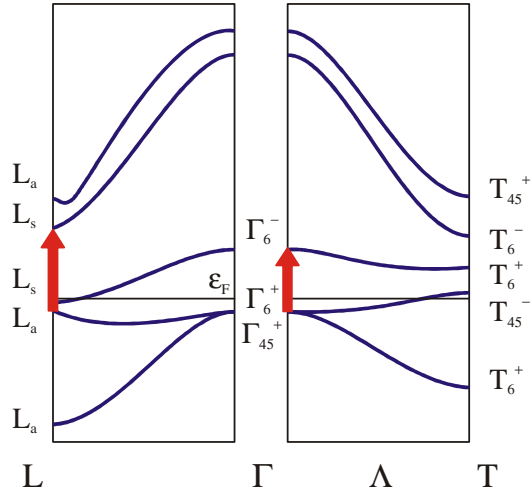

Fig. 5. A fragment of band structure of bismuth reproduced from [18]. Arrows indicate optical transitions with energies close to 1 eV (see text).

There is, however, a strong evidence against any general form of RM for fully symmetric mode. We compared the amplitudes of $A_{1g}$ and $E_g$ modes obtained in our time-resolved experiments with the corresponding Raman cross sections measured in the 1.55-2.7 eV range [20] and reproduced in Fig. 6. The comparison shows that only for $E_g$ mode the Raman cross section and the coherent phonon amplitude demonstrate similar behavior on excitation wavelength (here we have disregarded a longer tail in the short wavelength range). A limited excitation range used in the cited Raman study doesn't allow finding out the exact maximum position of this resonance profile. However, in the same study it was suggested that the most effective Raman scattering should take place for photon energies in the vicinity of $E_1$ critical point of bismuth (1.2 eV) [20]. This point corresponds to $L_a(5) - L_s(7)$ electronic transition mentioned above.

In contrast to $E_g$ oscillations, the dependence of fully symmetric $A_{1g}$ coherent amplitude on pump wavelength is essentially different from the resonance profile of $A_{1g}$ scattering cross section. Indeed, the coherent $A_{1g}$ phonons are generated effectively only when the energy of pump photons exceed a critical value that can be roughly defined as ~ 0.7 eV. If this condition is fulfilled, the amplitude, and even the whole photoinduced response (if we ignore $E_g$ oscillations and a small additive constant), demonstrate almost no dependence on excitation wavelength. This result is essentially different from that obtained in [10] with degenerate pump-probe technique for fully symmetric phonons in Sb and used there to support the unified Raman model.

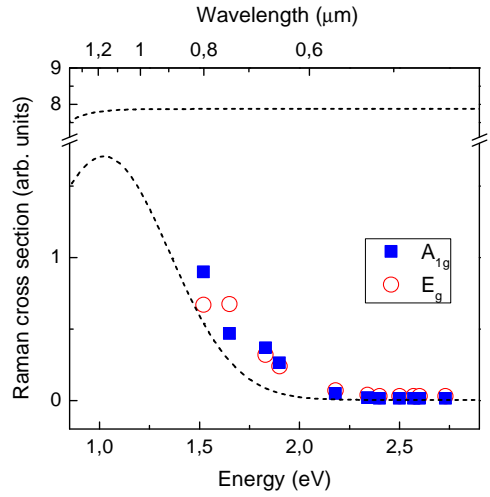

Fig. 6. Resonance profiles of the Raman cross sections of $E_g$ and $A_{1g}$ phonon modes of bismuth measured in [20]. Dashed lines are the same $A_{1g}$ and $E_g$ profiles as in Fig. 3, but normalized in a way to facilitate the comparison.

The observed discrepancy between the $A_{1g}$ resonance Raman profile and the pump wavelength dependence allows suggesting that RM in the case of fully symmetric phonons of bismuth is not effective (or its efficiency is significantly smaller than that of DECP). Indeed, if both $A_{1g}$ and $E_g$ modes were excited by the same Raman-like mechanism, the amplitude of higher frequency $A_{1g}$ mode would be reduced in comparison to the amplitude of low frequency $E_g$ mode (thus their amplitude ratio would be smaller than in the frequency domain). This is a direct consequence of different ratios between the pulse bandwidth and the phonon frequency that define the efficiency of coherent phonon generation: the larger the ratio, the more effective is the generation. However, in the experiment we observe the opposite situation: the amplitude ratio in the time domain is significantly larger than in the frequency domain.

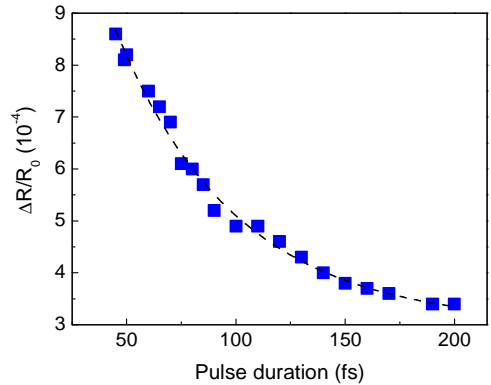

Fig. 7. Coherent $A_{1g}$ amplitude as a function of pump pulse duration. The dashed line is an exponential fit (see text). The data were obtained in a degenerate pump-probe scheme at $\lambda$=800 nm. The probe pulse was always transform limited one with duration of 45 fs.

To demonstrate the dependence of coherent amplitude on the bandwidth/frequency ratio, we measured it for fully symmetric phonons by stretching the pump pulse duration. The result shown in Fig. 7 indicates that the corresponding dependence is almost exponential when fitted to the function *exp(-t/τ)+const*, with a characteristic time $\tau$ of approximately one quarter of inverse phonon frequency. Thus, in the time-domain the amplitude ratio for $A_{1g}$ and $E_g$ modes should be around 1.5, instead of more than 10 observed at excitation wavelengths away from resonance at λ=1300 nm, where it is approximately 4.

In an attempt to explain the specific form of the wavelength dependence for $A_{1g}$ phonon amplitude we have analyzed linear optical properties of bismuth and found out that the coherent amplitude $A(A_{1g})$ correlates not with the resonance Raman profile but with the absorption spectrum of the crystal. To illustrate this we show in Fig. 8 the absorption coefficient *k* and the penetration depth *d* calculated from ellipsometric data [19]. It should be noted that the probe wavelength in all our measurements is fixed and therefore, the same volume is probed for each excitation wavelength. When the pump wavelength differs from λ=800 nm, which is our probe wavelength, the absorption length varies and so does the excited volume. Thus, to treat the effect of absorption on coherent phonon amplitude properly, one should compare $A(A_{1g})$ with penetration depth *d*. Indeed, as follows from the results presented in Fig. 3 and Fig. 8, in the range of 0.5-1 eV the fully symmetric coherent amplitude varies from ~ 0.003 to ~ 0.012, while the penetration depth *d* – from ~ 30 nm to ~ 110 nm: both demonstrate approximately four-fold increase. Such a similarity between the absorption and the coherent $A_{1g}$ amplitude suggests that the lattice state just after ultrafast excitation is primarily controlled by the amount of absorbed energy.

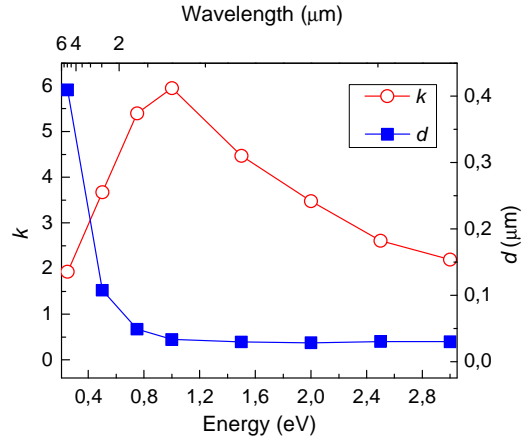

Fig. 8. Absorption coefficient *k* and penetration depth *d* of bismuth as a function of photon energy calculated using the data provided in [19].

The key role of the amount of absorbed energy and extraordinary high efficiency for fully symmetric coherent phonons naturally suggest that the generation mechanism in this case should be considered as purely displacive one. Indeed, rhombohedral bismuth has a Peierls-distorted lattice, which is highly sensitive to the state of electronic subsystem. Specifically, the increased number of excited electrons near the Fermi level can significantly alter interatomic potential. At the same time, the discussed $A_{1g}$ mode is unique as only its symmetry coincides with that of the Peierls shift [8]. We believe that the generation $A_{1g}$ coherent phonons involves the promotion of electrons to the $\Gamma_6^-$ point of the Brillouin zone (see Fig. 5) involving $\Gamma_{45}^+(4) - \Gamma_6^-(6)$ and $\Gamma_6^+(5) - \Gamma_6^-(6)$ optical transitions. It is reasonable as the Peierls distortion in bismuth occurs along the Γ-Λ-T line in the reciprocal space. Thus, the fully symmetric atomic oscillations can be launched by an ultrashort laser pulse if and only if a considerable part of pump pulse energy has been absorbed, creating a nonequilibrium electronic distribution. This process is completely different from the Raman excitation of coherent phonons. If in the Raman mechanism an ultrashort light pulse literally kicks the lattice, sending it into motion, and the momentum and energy transferred in this kick depend on the strength of the light pulse alone, in DECP the laser pulse kicks the potential and the energy conveyed to the lattice basically depends on excited charge density. RM is a *direct* two photon process, while DECP is a *two-step* process, which essentially depends on the relaxation of excited electron density. Their counterparts in the frequency domain are resonant Raman scattering and hot luminescence, respectively [21]. The major difference between the two time-domain processes is that RM relies on the transverse relaxation creating lattice coherence due to laser field coherence, whereas DECP is affected by the longitudinal relaxation and produces lattice coherence by a rapid nonradiative process. Our results suggest that in the case of $A_{1g}$ oscillations in bismuth this nonradiative process is electronic thermalization. Indeed, laser pulses with photon energy in the visible range produce nearly identical photoinduced responses and, in particular, fully symmetric coherent phonons of equal amplitude. Therefore, the coherent excited state is formed through the redistribution of energy of highly nonequilibrium charge carriers. Since thermalization is faster than inverse period of $A_{1g}$ oscillations, the generation of coherent atomic vibrations becomes possible.

## 5. Summary

In conclusion, we have measured the amplitude of coherent phonons of $A_{1g}$ and $E_g$ symmetry in bismuth in visible and near infrared ranges and found evidences against the unified Raman model. The comparison of the obtained data to spontaneous Raman cross sections and absorption spectrum allowed us to directly differentiate between DECP mechanism for $A_{1g}$ and Raman mechanism for $E_g$ phonon modes. While fully symmetric lattice dynamics is governed by the excited electronic density, the doubly degenerate phonon mode is "field driven" and demonstrates a distinct resonance behavior. The latter circumstance makes it possible to generate $E_g$ coherent phonons with about 1:4 ratio for $E_g$ to $A_{1g}$ amplitudes by using the pump wavelength of $\lambda$=1300 nm. Based on the observed differences of $E_g$ and $A_{1g}$ coherent modes, we conclude that displacive excitation, which requires a fast longitudinal relaxation, cannot be reduced in the case of Bi to Raman scattering based on transverse relaxation.

**Acknowledgements**

This work was partially supported by Russian Foundation for Basic Research.